# Evaluation of spatiotemporal tungsten density profiles using Unresolved Transition Arrays in the Large Helical Device

(submitted to Nuclear Fusion)

R. Nishimura[1,*], T. Oishi[1], I. Murakami[2,3], D. Kato[2,4], H. A. Sakaue[2], S. Gupta[5], H. Ohashi[6], C. Suzuki[2,3], M. Goto[2,3], Y. Kawamoto[2,3], R.T. Ishikawa[2,3], T. Kawate[7], K. Mukai[2,3], B.J. Peterson[2,3] and H. Takahashi[1]

[1] Department of Quantum Science and Energy Engineering, Tohoku University, 6-6-01-2 Aobayama, Sendai 980-8579, Japan
[2] National Institute for Fusion Science, National Institutes of Natural Sciences, 322-6, Oroshi-cho, Toki 509-5292, Japan
[3] Graduate Institute for Advanced Studies, SOKENDAI, 322-6, Oroshi-cho, Toki 509-5292, Japan
[4] Interdisciplinary Graduate School of Engineering and Sciences, Kyushu University, Kasuga 816-8580, Japan
[5] Institute of Space and Plasma Sciences, National Cheng Kung University, Tainan 70101, Taiwan
[6] Institute of Liberal Arts and Sciences, University of Toyama, Toyama 930-8555, Japan
[7] Naka Institute for Fusion Science and Technology, National Institutes for Quantum Science and Technology, 801-1 Mukoyama, Naka 311-0193, Japan

**E-mail:** ryota.nishimura.b8@tohoku.ac.jp

Abstract

Tungsten spectroscopic studies have been conducted in the Large Helical Device with a pellet injection technique. Spatiotemporal profiles of tungsten density were evaluated using a space-resolved spectrometer, for plasmas with an electron temperature of below 1 keV and electron density of $10^{19}$-$10^{20}$ $m^{-3}$. Slice & Stack, a method for reconstructing emissivity, was applied to a line at 191.7 Å, which is a part of the Unresolved Transition Array (UTA) spectrum at 90-250 Å. Tungsten density was obtained using photon emission coefficients of $W^{17+}$-$W^{27+}$, evaluated from collisional-radiative model. The tungsten pellet injected from outside the plasma was first ablated in the edge plasma and subsequently diffused throughout the plasma. This behavior is typical of pellet injection experiments. Furthermore, after an event triggered by NBI breakdown, tungsten accumulated in the core plasma. The radiation power was estimated from the evaluated tungsten density profile and cooling factor dataset, and compared with bolometer measurement. This sequence of processes would be useful for validating atomic data of tungsten ions in low-to-intermediate charge states.

## 1. Introduction

Tungsten (W, Z=74) is a candidate material for plasma facing components in the divertor and first wall of ITER and future demonstration fusion reactors [1,2], because of its optimal properties, such as high thermal conductivity, high melting point, and low sputtering yield. However, accumulation of tungsten impurities in the plasma would lead to the loss of a large amount of the plasma-stored energy by radiation, ionization, recombination, and excitation. Thus, the tungsten density is a key parameter that should be monitored in long-pulse operation.

Spectroscopic measurements of tungsten emission spectra have been conducted in many high-temperature plasma devices, such as ASDEX Upgrade [3], JET [4], JT-60U [5], EAST [6], WEST [7], HL-2A [8], and the Large Helical Device (LHD) [9]. A method for estimation of tungsten concentration has been established by Asmussen *et al* [10]. They observed tungsten Unresolved Transition Arrays (UTAs) around 50 Å, and estimated tungsten concentration by comparing the radiation power and the intensity of $W^{45+}$ emission using the modified ionization and recombination rate coefficients. In contrast to analyses using UTAs, Zhang *et al* evaluated tungsten influx rate using S/XB ratio, which denotes the number of ionization events per photon emission, of two $W^{5+}$ lines at 382.13 Å and 394.07 Å [11]. In addition to impurity diagnostics, electron temperature and density diagnostics have also been performed. Recently, Li *et al* reported Line-Intensity Ratio (LIR) data for open 4d shell ions, from $W^{31+}$ to $W^{39+}$, based on results of collisional-radiative model (CRM) [12]. A notable feature of their LIR data is the use of pairs of emission lines from ions in different charge states. Although experimental validation is still required, the LIR data could serve as an effective tool for plasma measurements.

Experimentally observed UTAs generally include several charge states. For tungsten, open 4d and 4f shell ions emit UTAs in vacuum ultraviolet (VUV) and extreme ultraviolet (EUV) regions. Several UTA spectra have been observed at 15–45 Å (4f–5g, 4f–6g, 4d–5p, 4d–5f, 4p–5s, 4d–5s) [13,14], 45–70 Å (n = 4–4 transitions, where n is principal quantum number) [15,16], 170–240 Å (n=5–5 transitions) [17], and 290–370 Å (5s–5p) [18–20]. In addition to spectral observations, evaluation of spatial profiles of tungsten ions has also been performed with a space–resolved EUV spectrometer using many lines of sight [21].

One of the key targets for space-resolved spectroscopy is local emissivity of tungsten ions. The local emissivity from $W^{q+}$, $\varepsilon_q$, is related to ion density, $n_q$, with the relationship $\varepsilon_q$ = PEC $n_e$ $n_q$, where PEC and $n_e$ are the photon emission coefficient and electron density. Furthermore, tungsten density, $n_W$, could be used to estimate radiation power density, $p_W$, with the relationship $p_W = L_W n_e n_W$, where $L_W$ is the cooling factor. Since the radiation power determines the power balance and control of fusion plasmas, evaluation of the spatiotemporal profiles of $n_W$ and $p_{rad}$ is particularly important. In this study, we focus on evaluation of the spatiotemporal profiles of tungsten density and radiation power in the tungsten pellet injection experiments using UTA spectrum around 200 Å.

## 2. Large Helical Device

### 2.1 Measurement systems

Top view of LHD [22] with visible, VUV and EUV spectrometers is shown in Fig. 1.

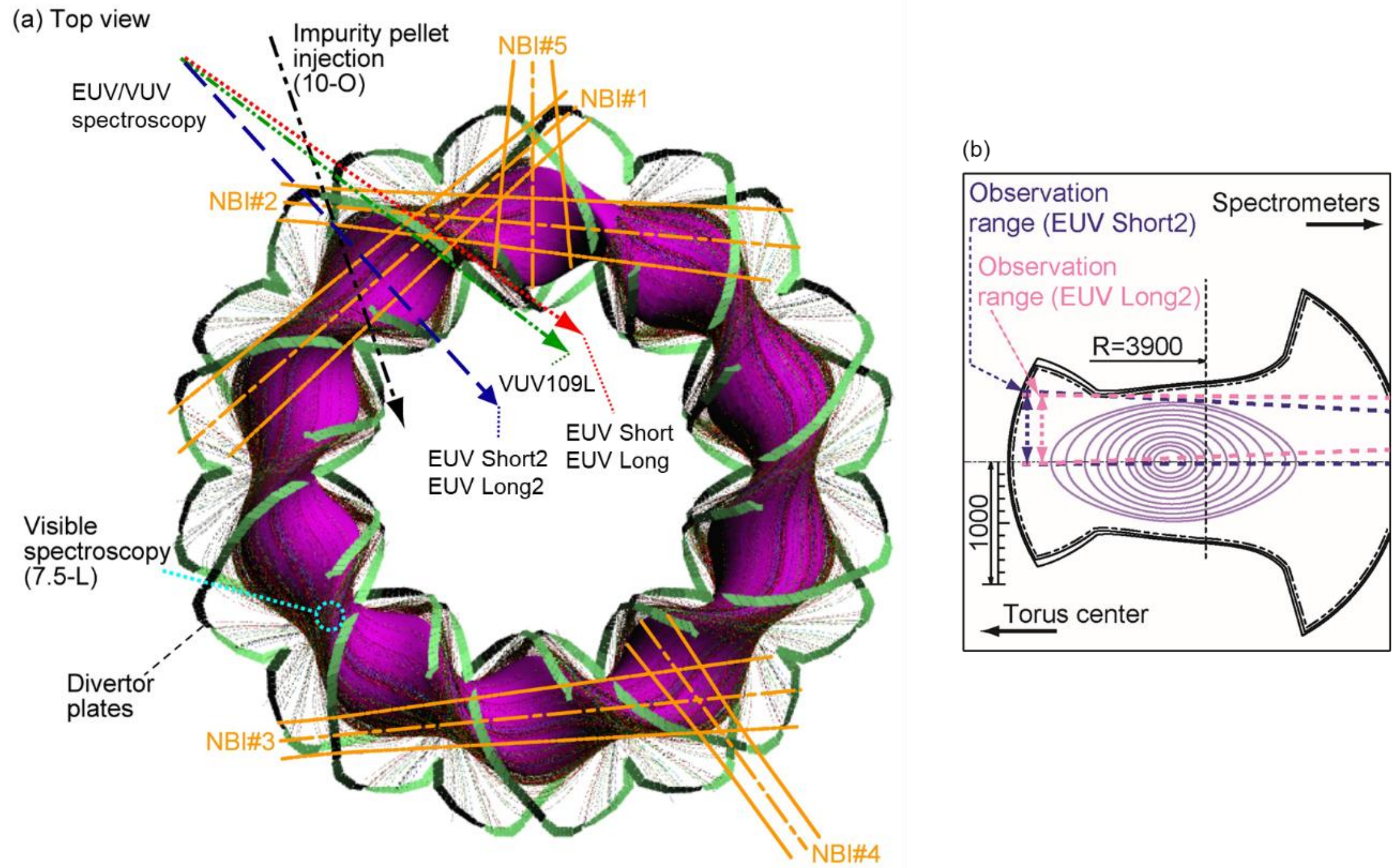

**Figure 1.** (a) Top view of LHD with schematic drawing of pellet injector and VUV/EUV spectrometers. (b) lines of sight of EUV Short2 and EUV Long2 space-resolved spectrometers.

A plasma is generated and sustained by electron cyclotron heating (ECH) and neutral beam injection (NBI) heating. A total of five NBI systems are installed, including three negative ion source NBIs (n–NBIs) and two positive ion source NBIs (p–NBIs). The p–NBIs are injected perpendicular to the magnetic axis, whereas n–NBIs are injected tangential to it.

Spectroscopic measurement systems, which cover the range from soft X-ray to visible, are installed in LHD. In this paper, a VUV spectrometer, called "VUV 109L" [23], and EUV spectrometers, called "EUV Short", "EUV Long", "EUV Short2" and "EUV Long2" [24], were used to observe tungsten spectra. The VUV 109L, EUV Short, and EUV Long spectrometers cover the wavelength ranges of 300–1050 Å, 5–130 Å, and 50–500 Å, respectively. The spectra are recorded at 5 ms intervals. The EUV Short2 and EUV Long2 are space-resolved spectrometers, which have a total 204 lines of sight. The spectra are taken at 100 ms intervals. In addition to spectroscopic measurements, a Thomson scattering system [25] and a bolometer [26] were used for diagnostics of electron temperature and electron density profiles, and radiation power, respectively.

Tungsten particles were intentionally supplied via tungsten pellet injection. A cylindrical tungsten pellet with a diameter of 0.05 mm and a length of 0.7 mm was used in this paper. Approximately $9\times10^{16}$ tungsten atoms were included in the pellet. The pellet injector is installed at the 10–O port. The time required for tungsten particles to become toroidally uniform can be estimated by comparing signals from bolometers installed at different ports. In many cases, injected tungsten particles diffuse uniformly in the toroidal direction in a very short time, typically less than 10 ms.

## 2.2 Tungsten spectral data collected in LHD

In LHD, more than 1,000 discharges with tungsten pellet injection have been conducted over the past decade. Emissions from up to $W^{46+}$, including UTAs, have been observed and identified (e.g., [27-30]). A summary of the observed tungsten spectra in LHD is shown in Fig. 2.

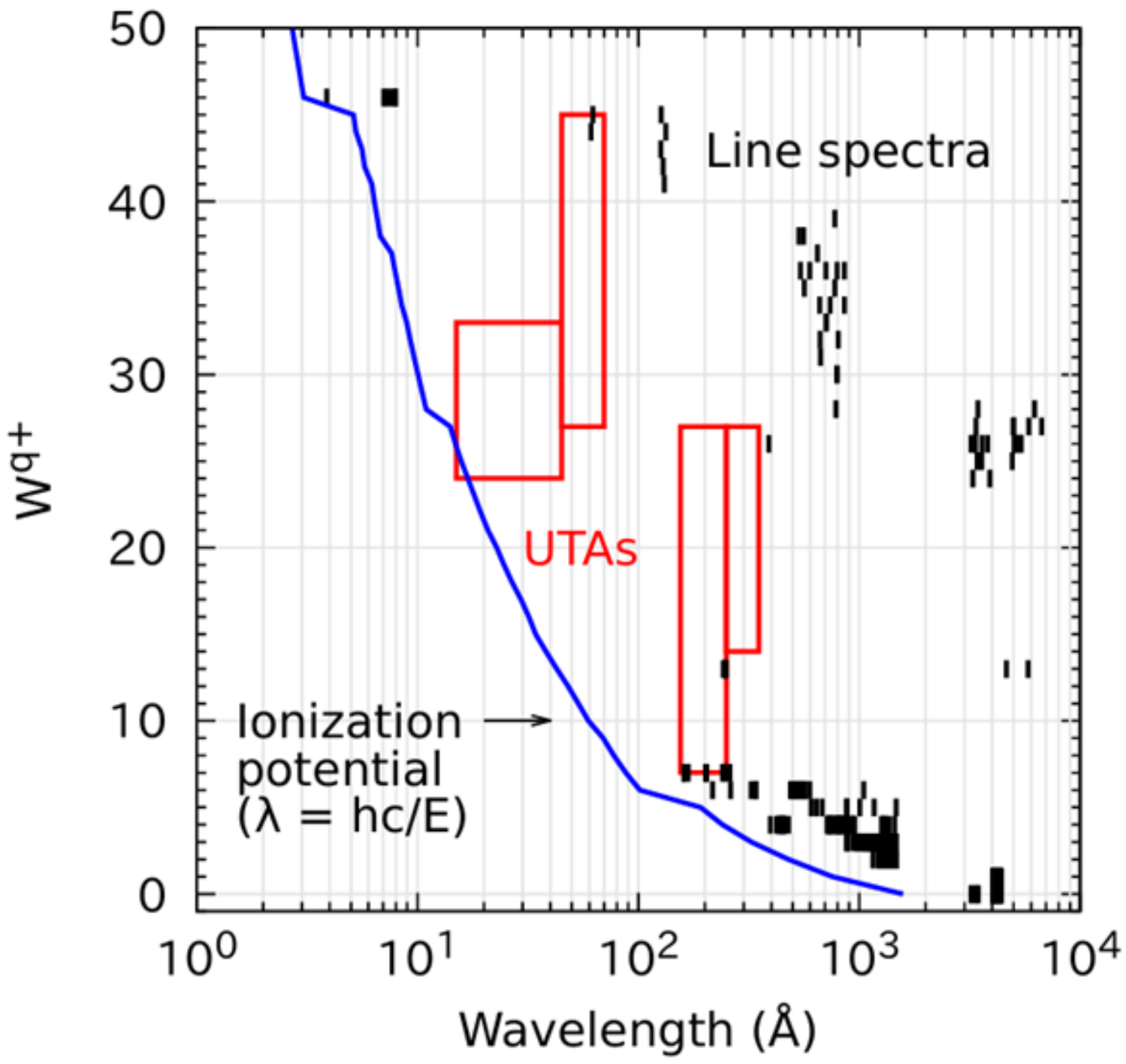

**Figure 2.** Tungsten spectra observed in LHD. Black vertical bars and red boxes show the observed line spectra and UTAs. The blue curve shows the ionization potential, E, plotted as the corresponding wavelength given by $\lambda = hc/E$, where h and c are Planck's constant and the speed of light, respectively.

Many spectral lines and arrays from tungsten ions, ranging from soft X-rays to the visible, have been observed in LHD. Among them, electric dipole (E1) transitions exhibit relatively strong intensities and have therefore been widely used for impurity diagnostics. In addition to E1 lines, higher order electric (E2, E3) and magnetic (M1, M2, M3) transitions are also observed. For instance, M1 lines from $W^{37+}$ at 646.7 Å, $W^{38+}$ at 532.9 and 559.0 Å, $W^{39+}$ at 774.8 Å, and $W^{41+}$ at 131.21 Å have been observed in LHD. In the case of $W^{41+}$, the corresponding E1 transitions occur at 5–8 Å ($n = 4 \rightarrow 3$) and 40–80 Å ($n = 4 \rightarrow 4$), whereas the M1 lines appear at longer wavelengths. This difference in wavelength ranges provides additional flexibility in impurity diagnostics. UTA spectra result from the overlap of emissions from ions in different charge states within a specific wavelength range. By applying a collisional-radiative model to appropriately separate the ion populations, these spectra can be used for impurity diagnostics. One of the key issues is to apply both line and UTA spectra together to impurity diagnostics.

## 3. Evaluation of spatiotemporal profiles of tungsten impurity

### 3.1 Tungsten pellet injection experiment

Figure 3 shows the temporal evolution of the tungsten pellet injection experiment. The port-through power of ECH and n–NBI heating is shown in figures 3(a). Temporal evolution of the central electron temperature, $T_{e0}$, central electron density, $n_{e0}$, radiation power, $P_{rad}$, and plasma stored energy, $W_P$, are shown in figures 3(b) – (e).

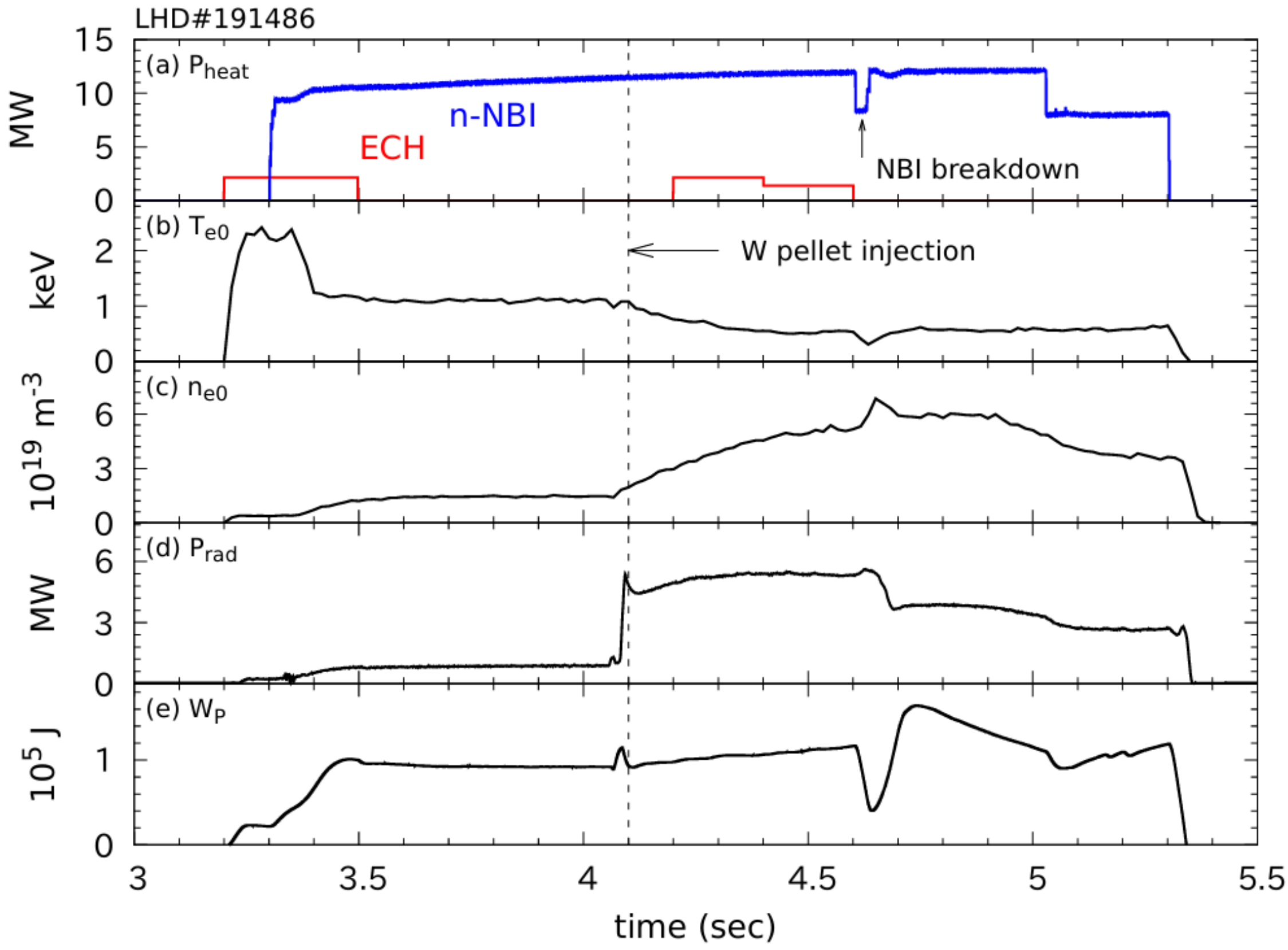

**Figure 3.** Temporal evolution of (a) port-through power of ECH and n-NBI heating, (b) $T_{e0}$, (c) $n_{e0}$, (d) $P_{rad}$, and (e) $W_P$ of a tungsten pellet injection experiment LHD#191486.

In this experiment, the magnetic axis and toroidal magnetic field strength were set at a major radius of $R_{ax}$ = 3.9 m and $B_t$ = 1.375 T, respectively. The plasma was ignited by ECH and maintained during about 2 s by n–NBI heating. During the phase in which the plasma is maintained by n–NBI, $T_{e0}$ remained below 1 keV. A tungsten pellet was injected into the plasma at $t$ = 4.1 s. Then, $P_{rad}$ abruptly increased and remained at a high level during the discharge. As tungsten particles were ionized, $n_{e0}$ gradually increased from $\mathbf{1.5 \times 10^{19}}$ m$^{-3}$ to $\mathbf{6 \times 10^{19}}$ m$^{-3}$. An event triggered by n-NBI breakdown was observed at $t$ = 4.65 s, and $W_P$ decreased instantaneously. The temporal evolution of $P_{rad}$ suggests that the tungsten particles accumulated in the plasma. Oishi *et al.* showed that tungsten particles accumulate when the electron density is higher than $\mathbf{2 \times 10^{19}}$ m$^{-3}$ [9]. Since the electron density at $t$ > 4.2 s remained above $\mathbf{2 \times 10^{19}}$ m$^{-3}$, it is consistent with their results.

### 3.2 **Emission spectra from tungsten ions at $T_e$ < 1 keV**

Temporal evolution of EUV and VUV spectra observed at different times are shown in Fig. 4. The spectra shown in figs. 4(a) – (b), (c) – (d), and (e) – (f) were observed using the EUV Short, EUV Long, and VUV 109L spectrometers, respectively.

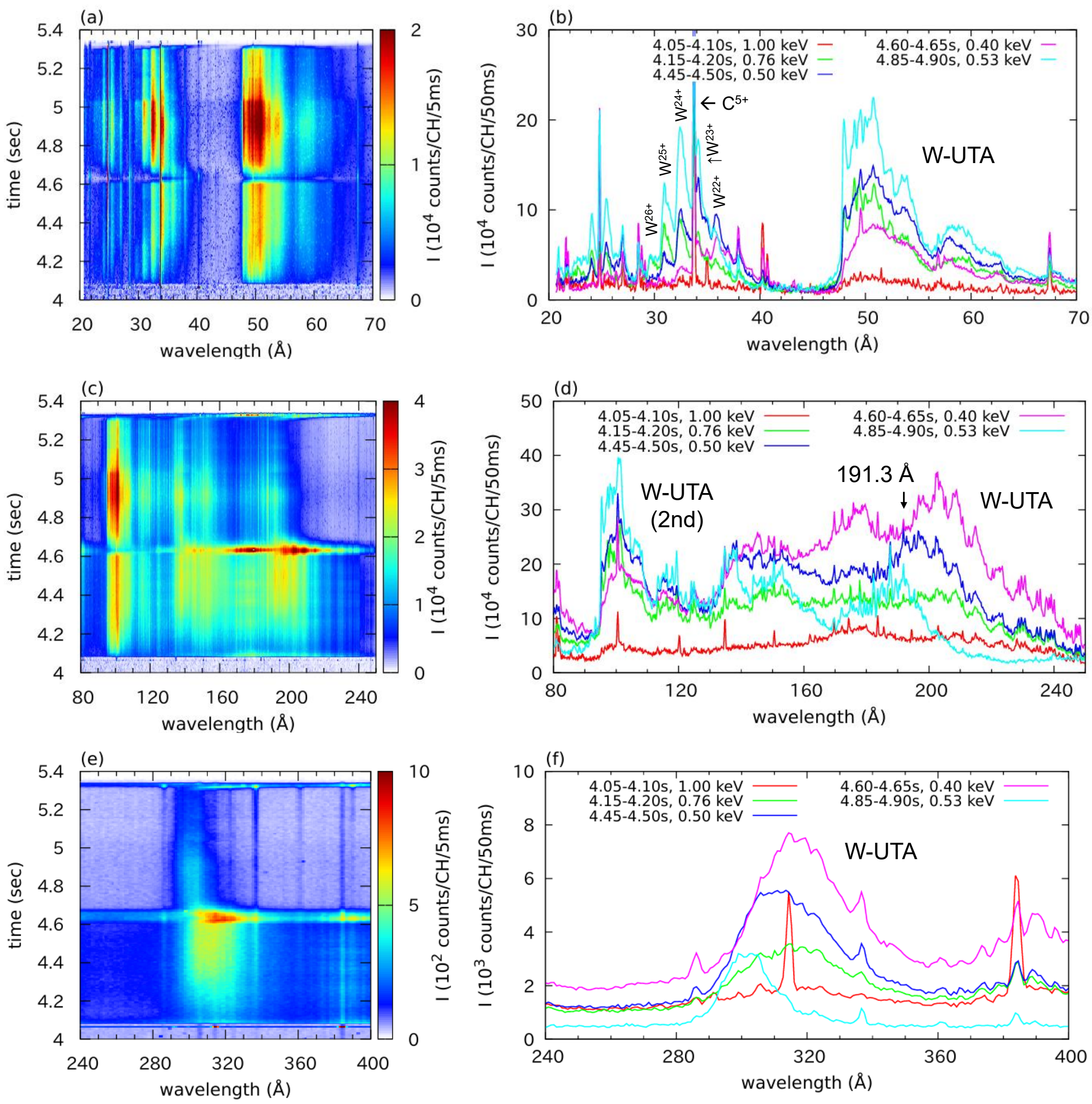

**Figure 4.** Measured emission spectra in LHD#191486. Spectra (a) – (b), (c) – (d), (e) – (f) were obtained using the EUV Short, EUV Long, and VUV 109L spectrometers, respectively. The legends in panels (b), (d), and (f) indicate the acquisition time-window and central electron temperature, $T_{e0}$. A tungsten pellet was injected at $t$ = 4.1 s.

Tungsten UTAs were clearly observed around 20–45 Å, 45–70 Å, 90–250 Å, and 280–350 Å. Each UTA spectrum shows different spectral shapes depending on $T_{e0}$.

$T_{e0}$ gradually decreased from 1 keV to 0.4 keV between 4.10 s and 4.65 s. Although $T_{e0}$ remained at a nearly constant value of 0.4 keV at $t > 4.65$ s, the emission intensity from ions in relatively high charge states, such as $W^{24+}$ (32.56 Å, 25.49 Å) and $W^{25+}$ (31.13 Å, 24.10 Å), increased. At this time, the UTA spectra around 200 Å and 300 Å shifted toward shorter wavelengths. This also indicates that the charge state shifted toward higher charge states. Since $T_{e0}$ exhibited only minor temporal variation at $t > 4.65$ s, the variation in spectral shape is primarily attributed to the spatial profiles of tungsten ions.

### 3.3 Collisional-radiative modelling for evaluation of photon emission coefficients

Spectral emissivity from $W^{q+}$, $\varepsilon_q$, is expressed as

$$\boldsymbol{\varepsilon_q = \mathrm{PEC}_q\, n_e n_q}\ , \quad (1)$$

where $\mathrm{PEC}_q$, $n_e$, and $n_q$ represent photon emission coefficient for $W^{q+}$, electron density, and ion density of $W^{q+}$, respectively. The PEC varies with the electron density and electron temperature, indicating the dependence of excitation and recombination processes on plasma conditions. A collisional-radiative (CR) model was used for evaluation of PECs in this study. In a CR model, the population density, $n_j$, is calculated as

$$\frac{dn_j}{dt} = \Gamma_{\mathrm{populate}} - \Gamma_{\mathrm{de-populate}}\ , \quad (2)$$

where $\Gamma_{\mathrm{populate}}$ and $\Gamma_{\mathrm{de\text{-}populate}}$ represent the total population and depopulation rates of level $\boldsymbol{i}$, including all relevant excitation, de-excitation, ionization, and recombination processes, expressed as

$$\Gamma_{\mathrm{populate}} = \sum_{i>j}\left(A_{ij} + C_{ij}^{d}\, n_e\right)n_i + \sum_{j>k} C_{kj}^{e}\, n_k\, n_e + C_j^{rr} n_{q+1}\, n_e\ , \quad (3)$$

$$\Gamma_{\mathrm{de-populate}} = \sum_{j>k}\left(A_{jk} + C_{jk}^{d}\, n_e\right)n_j + \sum_{i>j} C_{ji}^{e}\, n_j\, n_e + C_j^{i}\, n_j\, n_e\ , \quad (4)$$

where $A$, $C^d$, $C^e$, $C^i$ and $C^{rr}$, represent radiative transition rate, rate coefficients for electron impact excitation, de-excitation, ionization, and radiative recombination. *i*, *j*, and *k* represent excitation levels. To solve Eq. (2), a quasi-steady state condition, $dn_j/dt = 0$, was assumed. In this study, Flexible Atomic Code [31] was used.

We improved upon the CR model presented in Ref. [17] by including higher excited states up to $n = 6$. The included electron configurations are $4f^m$, $4f^{m-1}$ 5l (l < 5), $4f^{m-1}$ 6l (l < 6), $4d^9\, 4f^{m+1}$, $4d^9\, 4f^m$ 5s, $4f^{m-2}\, 5s^2$, $4f^{m-2}$ 5s 5p, $4f^{m-2}\, 5p^2$, $4f^{m-2}$ 5s 5d, $4f^{m-1}$, $4d^9\, 4f^m$, and $4f^{m-1}$ 5l (l < 5), where m = 28-q (q=17-27). For $W^{27+}$ ([Kr] $4d^{10}$ 4f), $4f^{m-2}\, 5s^2$, $4f^{m-2}$ 5s 5p, $4f^{m-2}\, 5p^2$, $4f^{m-2}$ 5s 5d configurations were excluded. First, the population densities were calculated for given electron temperature and electron density. The spectra for $W^{17+}$-$W^{27+}$ were then weighted by the corresponding ion abundances and summed to obtain the total spectrum. The rates of ionization and recombination were obtained from ADF-11 file ("scd50_w.dat" for ionization rate and "acd50_w.dat" for recombination rate) on OPEN-ADAS [32]. The line spectra of $W^{17+}$-$W^{27+}$ were involved assuming Gaussian profiles with a full width at half maximum (FWHM) of 0.2 Å. The observed spectra and evaluated PEC spectra are shown in Fig. 5.

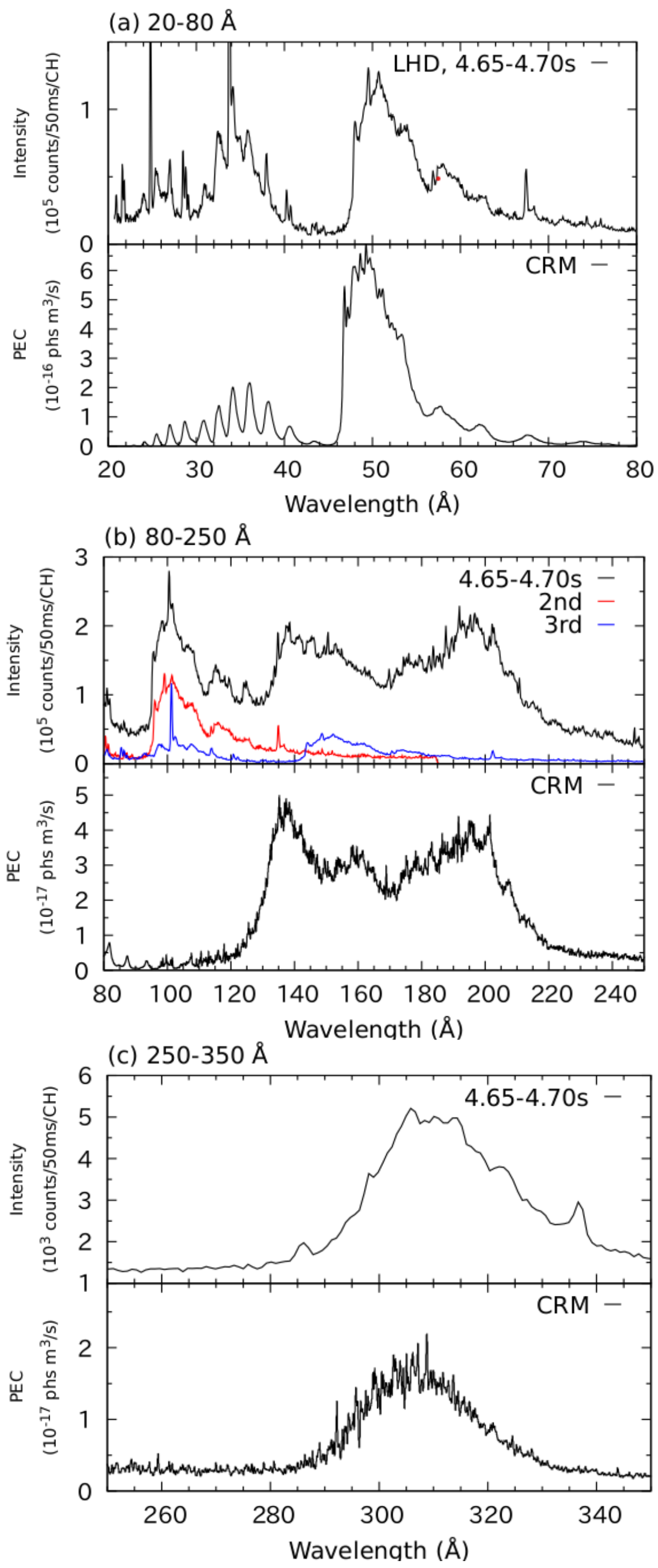

**Figure 5.** Comparison of the observed spectrum with the spectrum synthesized using the CR model. Panels (a) – (c) show the wavelength ranges in 20–80 Å, 80–250 Å, and 250–350 Å, respectively. Plasma condition with electron temperature of 0.5 keV and electron density of $10^{19}$ $m^{-3}$ was assumed for the CR model.

For the experimentally observed spectra shown in the upper panel of Fig. 5(b), the contributions from the second- and third-order diffraction are also shown. Despite the significantly more complex structure of UTAs, the spectra reproduced by the CR model show good agreement with the observed spectra. The observed spectra shown in the upper panels of Figures 5(a)–(c) represent line-integrated intensities, whereas the lower panels show local values. This suggests that both the line intensity ratios and the charge state distribution (CSD) are slightly influenced by line integration effects.

In the discharge #191486, emission intensity in the wavelength range of 200-240 Å decreased after $t$ = 4.65 s. The EUV Long2 spectrometer measured the wavelength range of 180-240 Å. For analyzing the spatial profile of tungsten, it is desirable to focus on spectral lines with relatively high emission intensity. For example, the lines at 184.6 Å, 191.7 Å, and 196.7 Å remained relatively high and stable intensity until the end of the discharge. Although any of these lines can be used to analyze the tungsten density profile, the line at 191.7±0.2 Å was used because of its particularly high emission intensity. The line at 191.7 Å is highlighted in Fig. 4(d).

The emission intensity at 191.7± 0.2 Å includes contributions from multiple charge states. PECs of $W^{17+}$–$W^{25+}$ as a function of electron temperature are shown in Fig. 6.

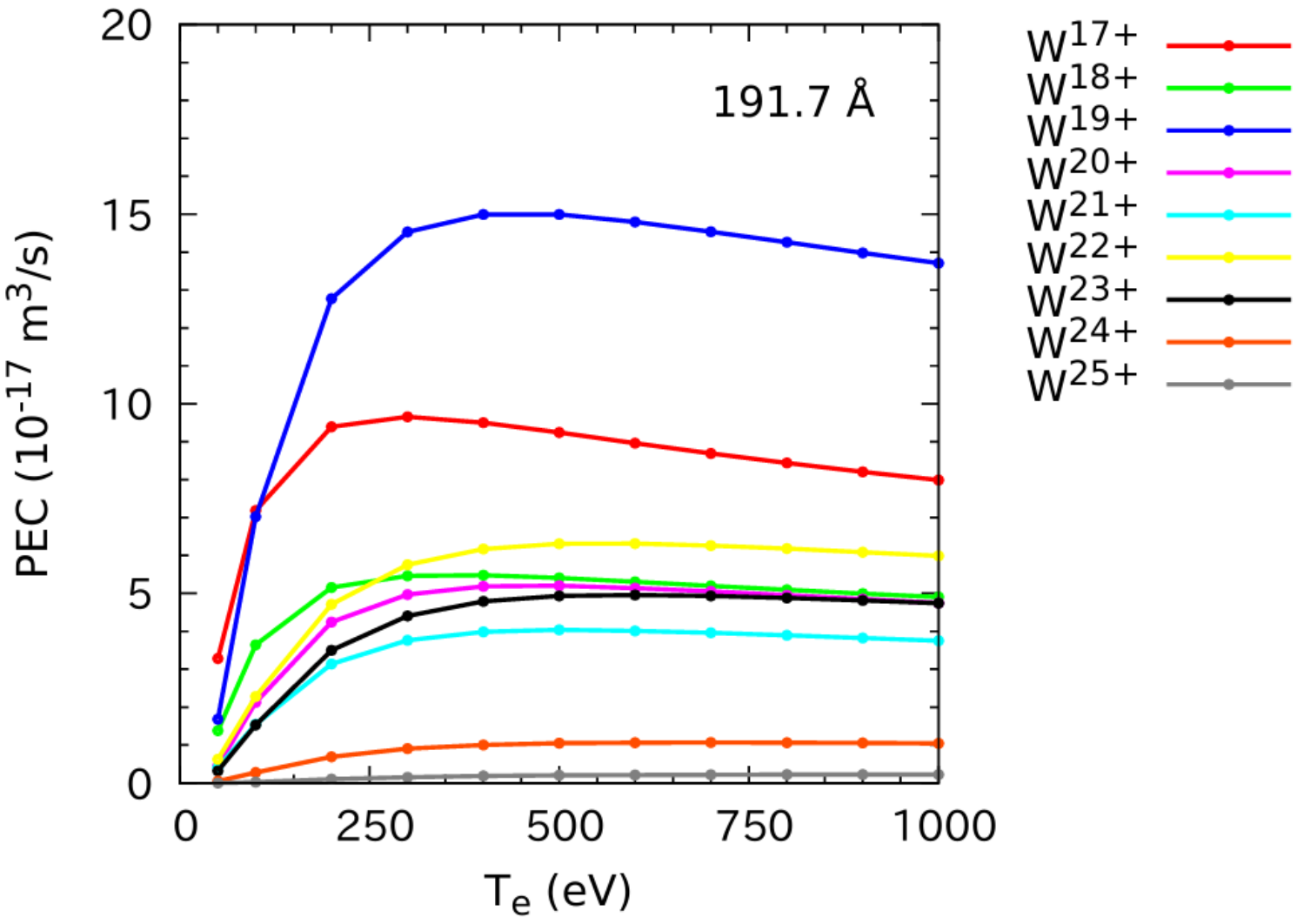

**Figure 6.** Photon emission coefficients averaged in 191.7± 0.2 Å of $W^{17+}$-$W^{25+}$ as a function of electron temperature. The electron density was assumed to be $10^{19}$ m$^{-3}$.

For the PECs of 191.7± 0.2 Å, $W^{19+}$ (5p–5d) is dominant, although contributions from other charge states are relatively small. Note that the PECs of $W^{26+}$ and $W^{27+}$ were negligibly small.

### 3.4 Evaluation of emissivity profiles

Tungsten ion density can be evaluated using PECs and local emissivity. The local emissivity was evaluated by the Slice & Stack method [21] using space-resolved spectroscopy. In the Slice & Stack method, line-integrated intensity, $I$, is expressed as

$$\boldsymbol{I} = \int_{LOS} \boldsymbol{\varepsilon}\, \boldsymbol{dL}\,, \qquad (5)$$

where d$L$ is length component along line of sight (LOS). The EUV Long2 spectrometer has a total of $\boldsymbol{N} = \mathbf{204}$ lines of sight. Thus, the following system of equations holds:

$$\boldsymbol{I} = \boldsymbol{L\varepsilon} \quad \rightarrow \quad \begin{bmatrix} \boldsymbol{I_1} \\ \boldsymbol{I_2} \\ \vdots \\ \boldsymbol{I_N} \end{bmatrix} = \begin{bmatrix} \boldsymbol{L_{11}} & \mathbf{0} & \mathbf{0} & \cdots & \mathbf{0} \\ \boldsymbol{L_{21}} & \boldsymbol{L_{22}} & \mathbf{0} & \cdots & \mathbf{0} \\ \vdots & \vdots & \vdots & \ddots & \vdots \\ \boldsymbol{L_{N1}} & \boldsymbol{L_{N2}} & \boldsymbol{L_{N3}} & \cdots & \boldsymbol{L_{NN}} \end{bmatrix} \begin{bmatrix} \boldsymbol{\varepsilon_1} \\ \boldsymbol{\varepsilon_2} \\ \vdots \\ \boldsymbol{\varepsilon_N} \end{bmatrix}. \quad (6)$$

Here, $L_{ij}$ denotes the path length of line of sight $i$ through magnetic flux surface $j$. The spatial distribution of $\boldsymbol{\varepsilon}$ is given by $\boldsymbol{\varepsilon} = \boldsymbol{L}^{-1}\boldsymbol{I}$. Since Eq. (6) involves a lower triangular matrix, it can be solved directly using forward substitution. In this case, the uncertainty in the emissivity at the plasma center becomes significant. By binning several lines of sight, the uncertainty in the emissivity can be reduced, while the spatial resolution is degraded. A typical spatial profile of emissivity, obtained by grouping six lines of sight and measured at 34 spatial points, is shown in Fig. 7(a). The sensitivity calibration of the EUV Long2 spectrometer was performed based on Ref. [33]. Figure 7(b) shows the observed line-integrated intensity $I$ together with the line-integrated intensity reproduced by the calculation ($\boldsymbol{I} = \boldsymbol{L\varepsilon}$).

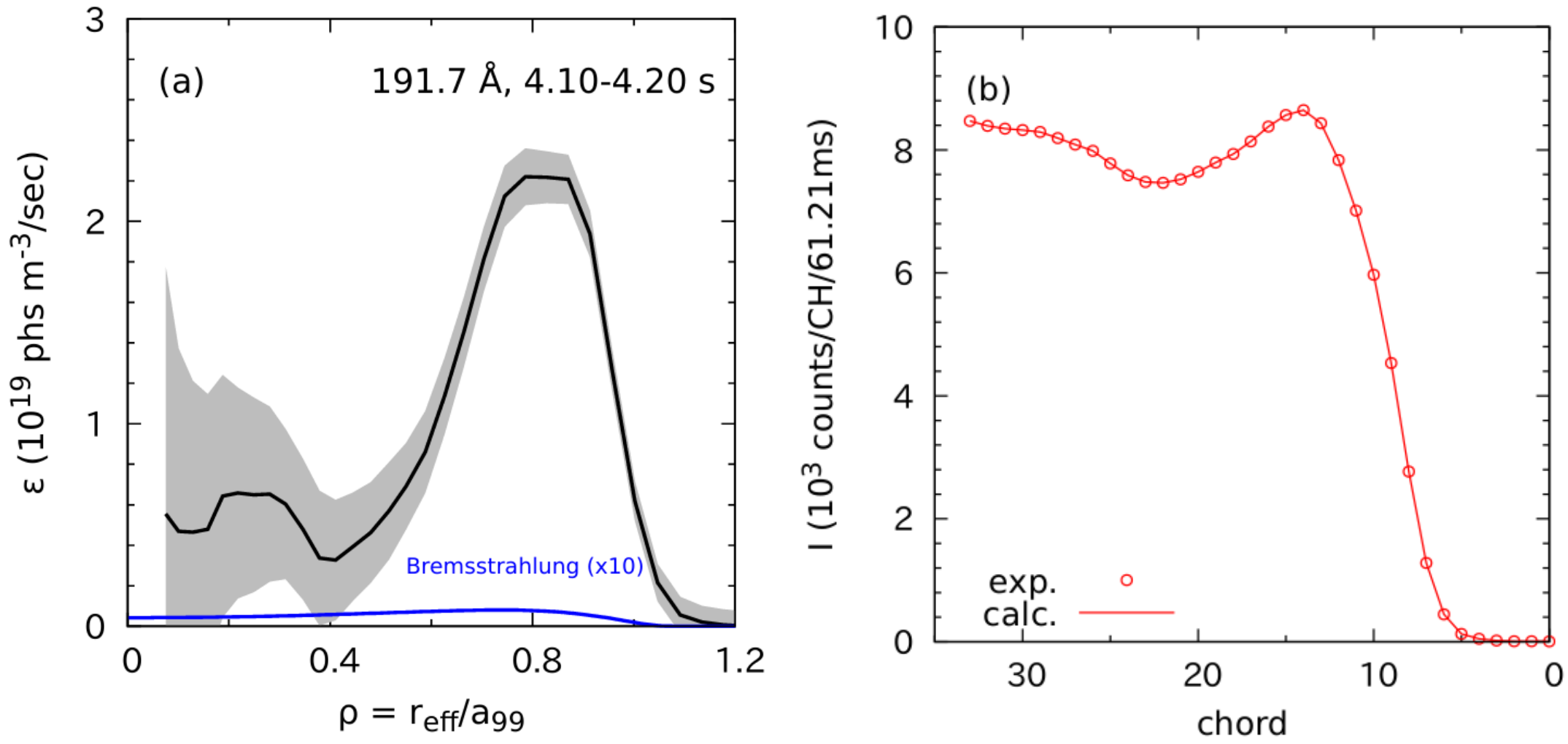

**Figure 7.** (a) Spatial profile of emissivity of 191.7±0.2 Å and bremsstrahlung, measured at $t$ = 4.10-4.20 s. (b) Comparison of the observed line-integrated intensity $I$ and that reproduced by the calculation ($\boldsymbol{I} = \boldsymbol{L\varepsilon}$).

For the emissivity profile of 191.7 ± 0.2 Å , the uncertainty is small at $\rho > 0.5$, where the emissivity is relatively high. In the core plasma ($\rho < 0.4$), the uncertainty is reduced by binning the lines of sight, remaining within a few times the local emissivity. The calculated emissivity reproduces the observed line-integrated intensity accurately. Note that the estimated bremsstrahlung emissivity profile was negligibly small. Accordingly, the emissivity profile can be employed for analyzing the tungsten density. Figure 8 shows the temporal evolution of the emissivity profiles of 191.7±0.2 Å from $t$ = 4.10 to 5.20 s.

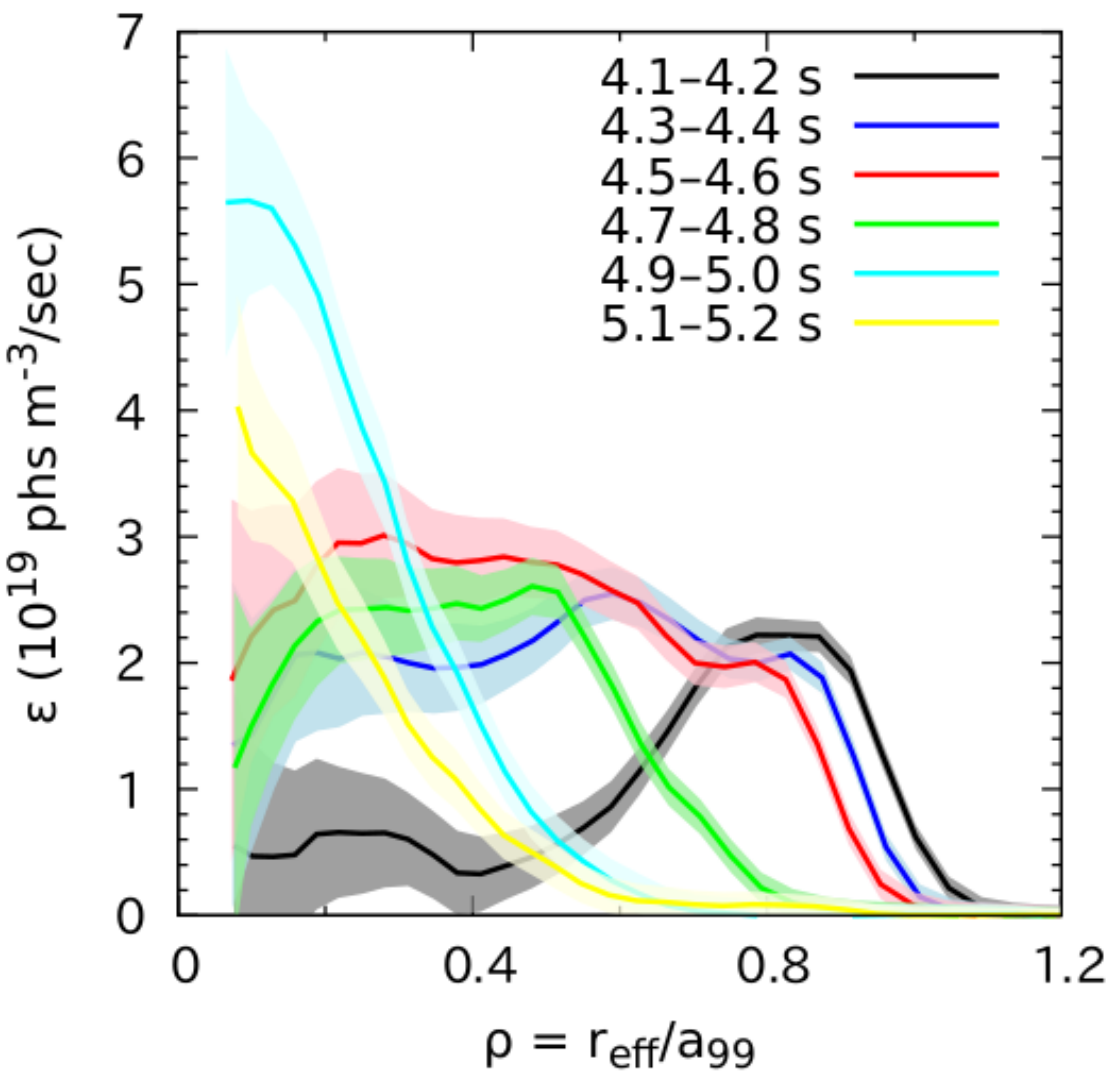

**Figure 8.** Temporal evolution of emissivity profiles of 191.7±0.2 Å at $t$ = 4.10-5.20 s.

Immediately after the pellet injection, at $t$ = 4.10-4.20 s, the emissivity of 191.7±0.2 Å at the plasma edge ($\rho$=0.8) is high. As time progresses, the emissivity in the core plasma gradually increases, and from $t$ = 4.9 s, the profile becomes localized in $\rho <$ 0.4. This result indicates that the tungsten was accumulated in the core plasma.

### 3.5 Evaluation of tungsten density profiles

As shown in Fig. 6, at a given wavelength in the UTA spectrum, emission includes contributions from many ions. Equation (1) can be written as follows:

$$\varepsilon = \sum_q \varepsilon_q = n_e \sum_q \mathrm{PEC}_q\, n_q . \quad (7)$$

Since Eq. (7) contains multiple unknowns, $n_q$, it is difficult to determine the ion densities. On the other hand, it has long been established, based on experiments on the TFR tokamak, that high-Z impurities such as tungsten ions are in ionization equilibrium within the existing uncertainties of the ionization and dielectronic recombination rate coefficients [34]. The ion density profile is determined by the ratio of ionization and recombination rates [35]. In LHD, it has also been shown that tungsten ions are in ionization equilibrium, as reported by Fujii *et al.* [36].

Assuming ionization equilibrium, $\boldsymbol{n_q = n_W f_q}$ , where $\boldsymbol{f_q}$ denotes the ion abundance of $W^{q+}$ and depends only on electron temperature and electron density, Eq. (7) can be rewritten as

$$\varepsilon = n_e\, n_W \sum_q \mathrm{PEC_q}\, f_q\, . \quad (8)$$

The quantity $\boldsymbol{f_q}$ is calculated from the electron temperature and electron density measured by a Thomson scattering system, together with the ionization and recombination rates provided in the OPEN-ADAS ADF-11 database [32]. These procedures allow the tungsten density to be evaluated.

A typical result of tungsten density evaluation is shown in Fig. 9.

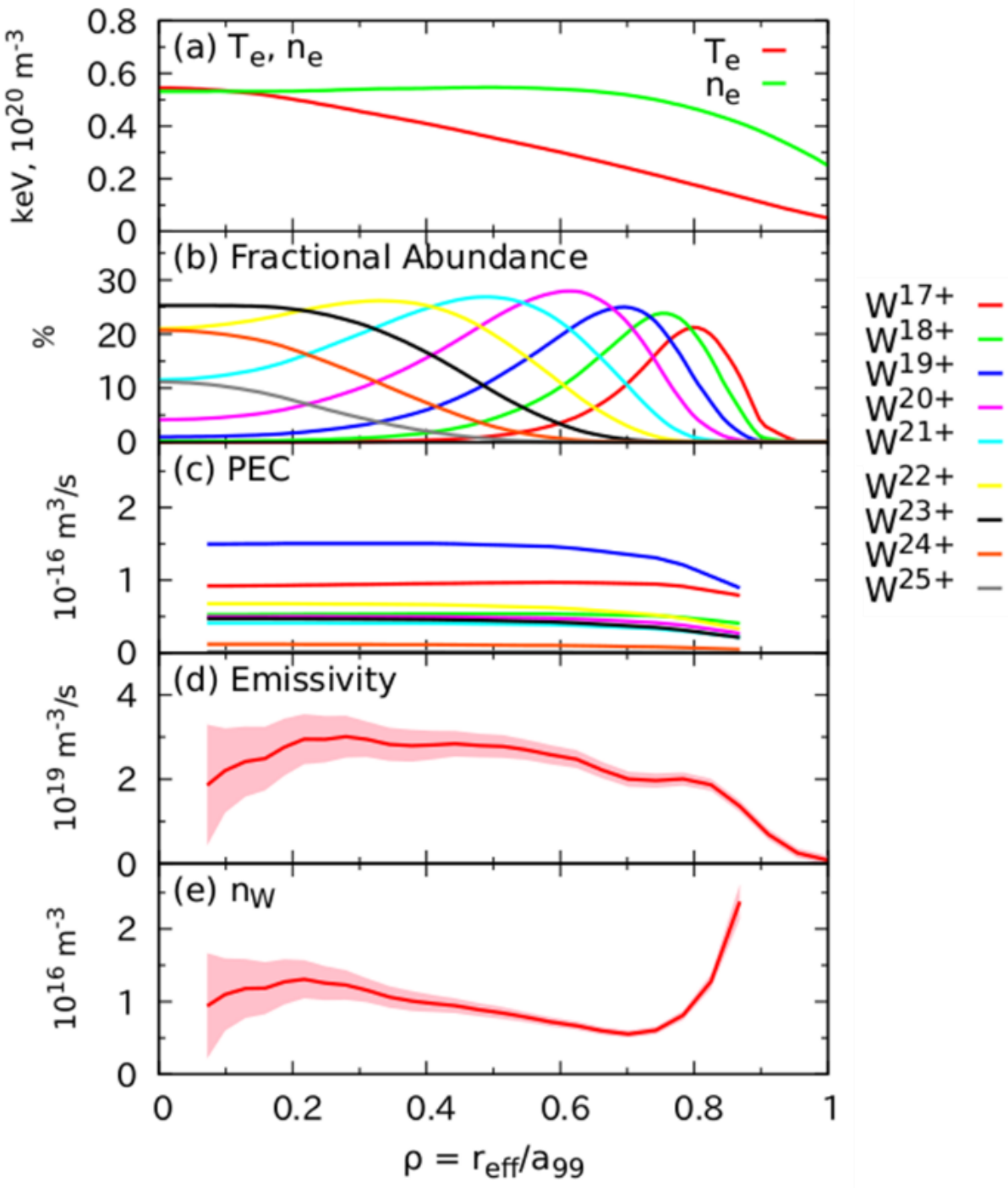

**Figure 9.** Typical result of tungsten density evaluation. Panels (a)-(e) represent the spatial profiles of $T_e$ and $n_e$, ion fractional abundances, PECs, Emissivity of 191.7±0.2 Å, and tungsten density at $t$ = 4.50-4.60 s.

In Eq. (8), when the abundances of all charge states from $W^{17+}$ to $W^{27+}$ are zero, the calculation of $\boldsymbol{n_W}$ breaks down. To ensure reliable results, we restrict the analysis to regions where $\boldsymbol{f_q}$ > 10 %. For an electron density of $10^{19}$ $m^{-3}$, the analyzable electron temperature range is 130–1200 eV. The absence of data for ρ > 0.9 in Fig. 9 is due to the temperature being below 130 eV.

At $t$ = 4.50-4.60 s, the evaluated $\boldsymbol{n_W}$ showed a peak at ρ > 0.8. In addition, $\boldsymbol{n_W}$ slightly increased from ρ = 0.1 toward ρ = 0.7. Further analysis is needed to understand this spatial profile of $\boldsymbol{n_W}$.

In pellet injection experiments, ablation first occurs in the edge plasma immediately after the pellet injection. Subsequently, the spatial profile is shaped by impurity transport. Therefore, it is necessary to determine the spatiotemporal profile. The spatiotemporal profiles of tungsten density are shown in Fig. 10.

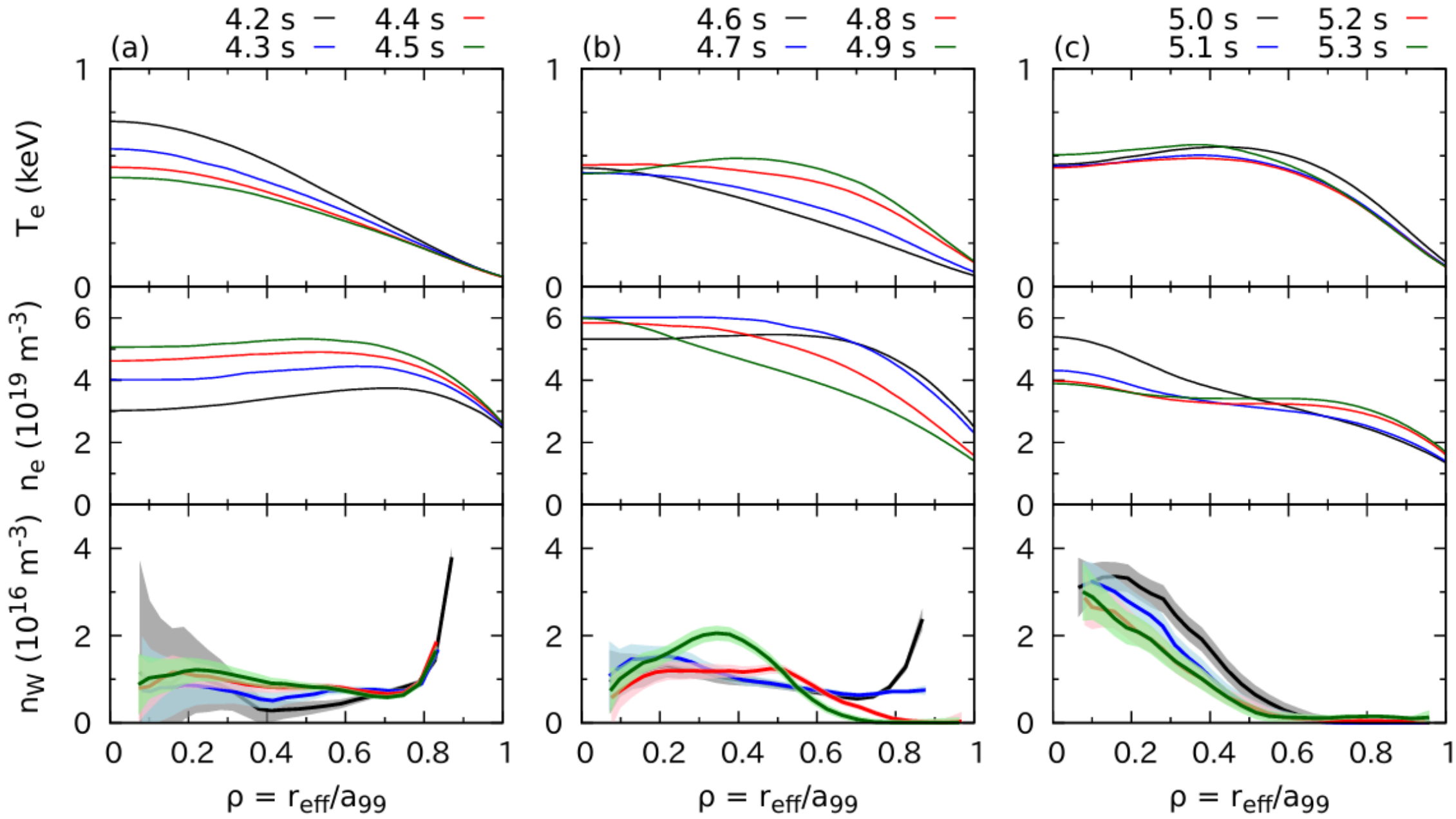

**Figure 10.** Spatiotemporal evolution of electron temperature, electron density, and tungsten density. Panels (a)-(c) show the time windows 4.20-4.50 s, 4.60-4.90 s, and 5.00-5.30 s, respectively.

After the pellet injection at $t$ = 4.10 s, tungsten was ablated around ρ = 0.9. The tungsten density near ρ = 0.9 remained high until $t$ = 4.60 s. After $t$ = 4.60 s, tungsten was clearly observed to accumulate from ρ ~ 0.9 toward ρ = 0 until $t$ = 5.30 s. For $t$ > 5.00 s, the tungsten density at the center of the plasma reaches $3 \times 10^{16}$ $m^{-3}$, corresponding to a relatively high tungsten concentration ($n_W/n_e$) of about $10^{-3}$.

The electron temperature profile becomes flattened within ρ < 0.6 after the pellet injection, whereas the electron density exhibits a centrally peaked profile. These features are consistent with the tungsten density profile. The observed accumulation of tungsten in the core plasma can be attributed to typical inward convective transport [29], although a detailed analysis is beyond the scope of this study.

As mentioned in Sec. 2, the number of injected particles, $N_W$, was about $9 \times 10^{16}$. The total number of tungsten particles can be obtained by volume integration of the tungsten density profile as

$$N_W = \int_V n_W(V)dV \,. \quad (9)$$

However, in Fig. 10, the tungsten density is given as a function of ρ, i.e., $n_W(\rho)$, and thus a variable transformation is required. In LHD, $dV/d\rho(\rho)$ has been measured by a Thomson scattering system. The profile of $dV/d\rho(\rho)$ at $t$ = 4.60 s is shown in Fig. 11.

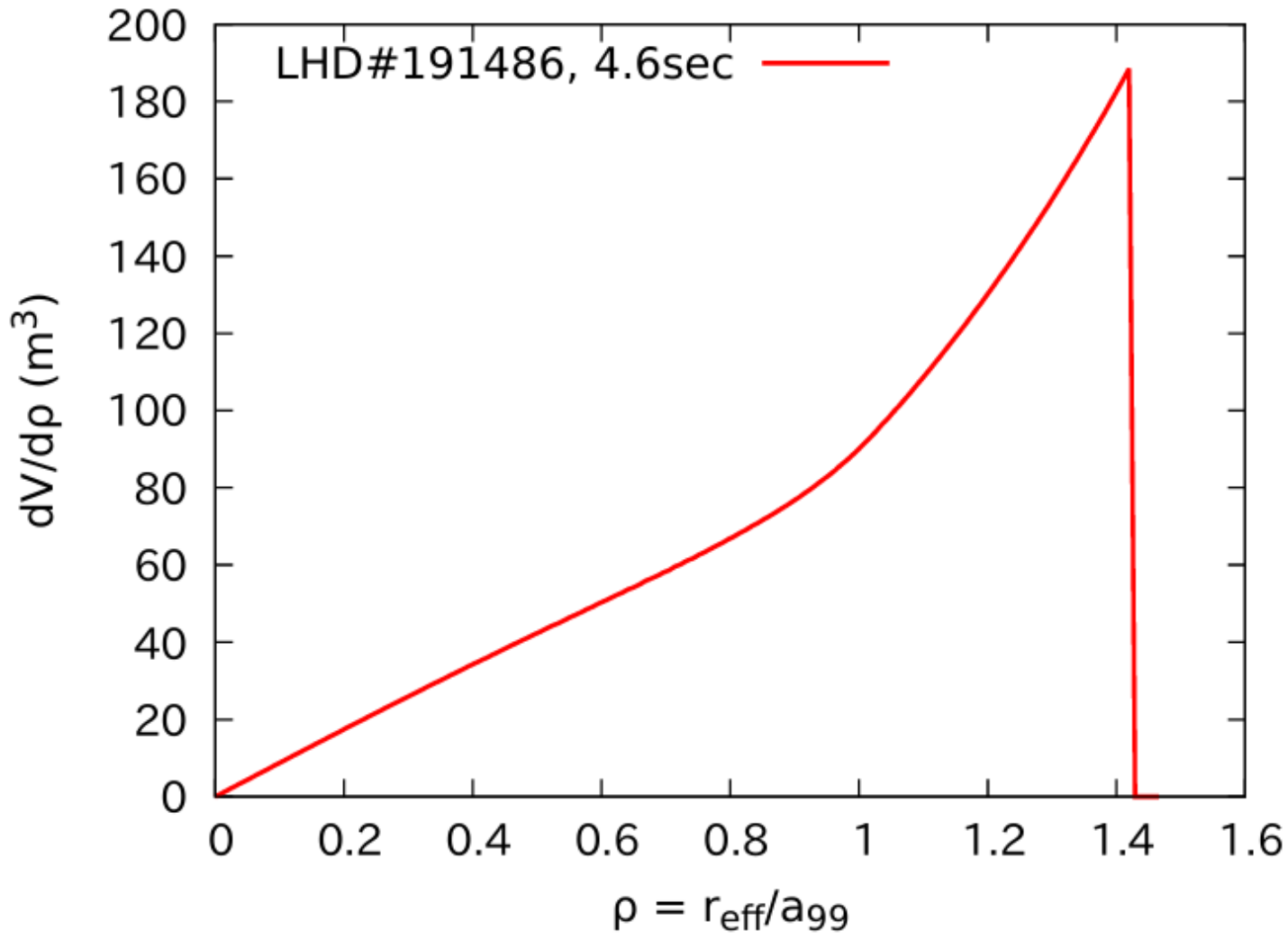

**Figure 11.** $dV/d\rho(\rho)$ profile, measured at $t$ = 4.60 s.

Using $dV/d\rho(\rho)$, Eq. (9) can be rewritten as

$$N_W = \int n_W(\rho)\frac{dV}{d\rho}(\rho)d\rho\,. \quad (10)$$

The temporal evolution of $N_W(t)$ evaluated based on this expression is shown in Fig. 12.

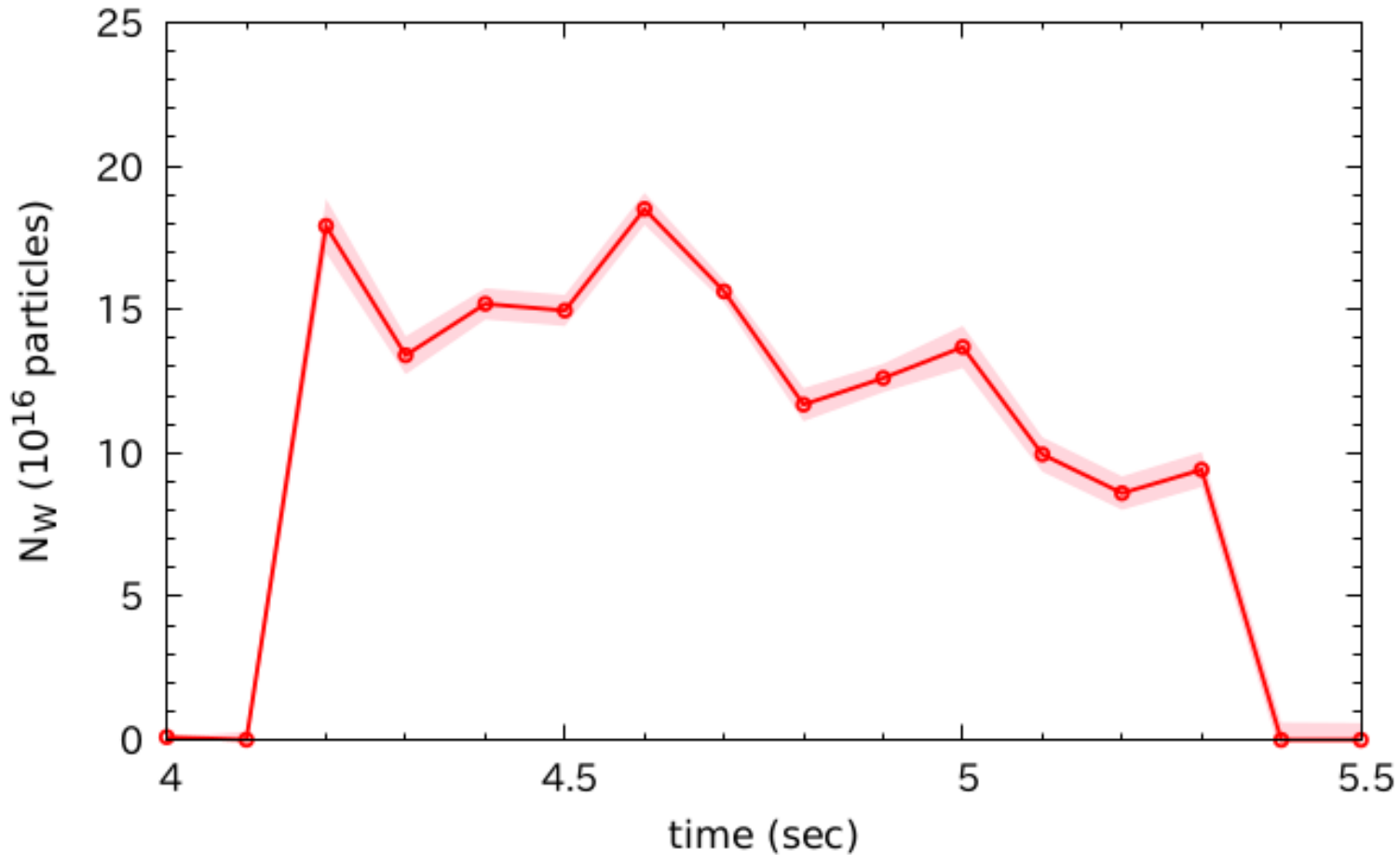

**Figure 12.** Estimated number of the tungsten particles exists in the plasma. The red region represents the ±1σ uncertainty.

$N_W(t)$ increased abruptly at $t$ = 4.20 s, and then gradually decreased until $t$ = 5.40 s. The maximum value of $N_W(t)$ was approximately $2 \times 10^{17}$, which is comparable to the number of injected particles of approximately $9 \times 10^{16}$.

### 3.6 Radiation power

One of the key objectives in tungsten spectroscopy is the evaluation of radiation power. The fundamental expression for the radiation power density of tungsten is given by $p_{\mathrm{W}} = n_{\mathrm{e}}\, n_{\mathrm{W}}\, L_{\mathrm{W}}$, where $L_{\mathrm{W}}$ represents the cooling factor. Several datasets of cooling factors have been reported in the literature. The ADAS dataset has been widely used to estimate radiation power. However, Peyrusse *et al.* suggested that the OPEN-ADAS data [32] may not be optimized for $T_{\mathrm{e}}$ < 1-2 keV [37]. Indeed, their results indicate that the cooling factor $L_W$ can be several times larger than the OPEN-ADAS values in this regime.

The radiation power is evaluated using an expression with the same form as Eq. (10):

$$P_{\mathrm{W}} = \int p_W(\rho) \frac{dV}{d\rho}(\rho)\, d\rho. \quad (11)$$

Figure 13 shows the temporal evolution of the radiation power evaluated using cooling factors, together with the radiation power measured by the bolometer. The radiation power was evaluated using cooling factor datasets provided by OPEN-ADAS and from Peyrusse *et al.*, and the results are compared with the bolometer measurements.

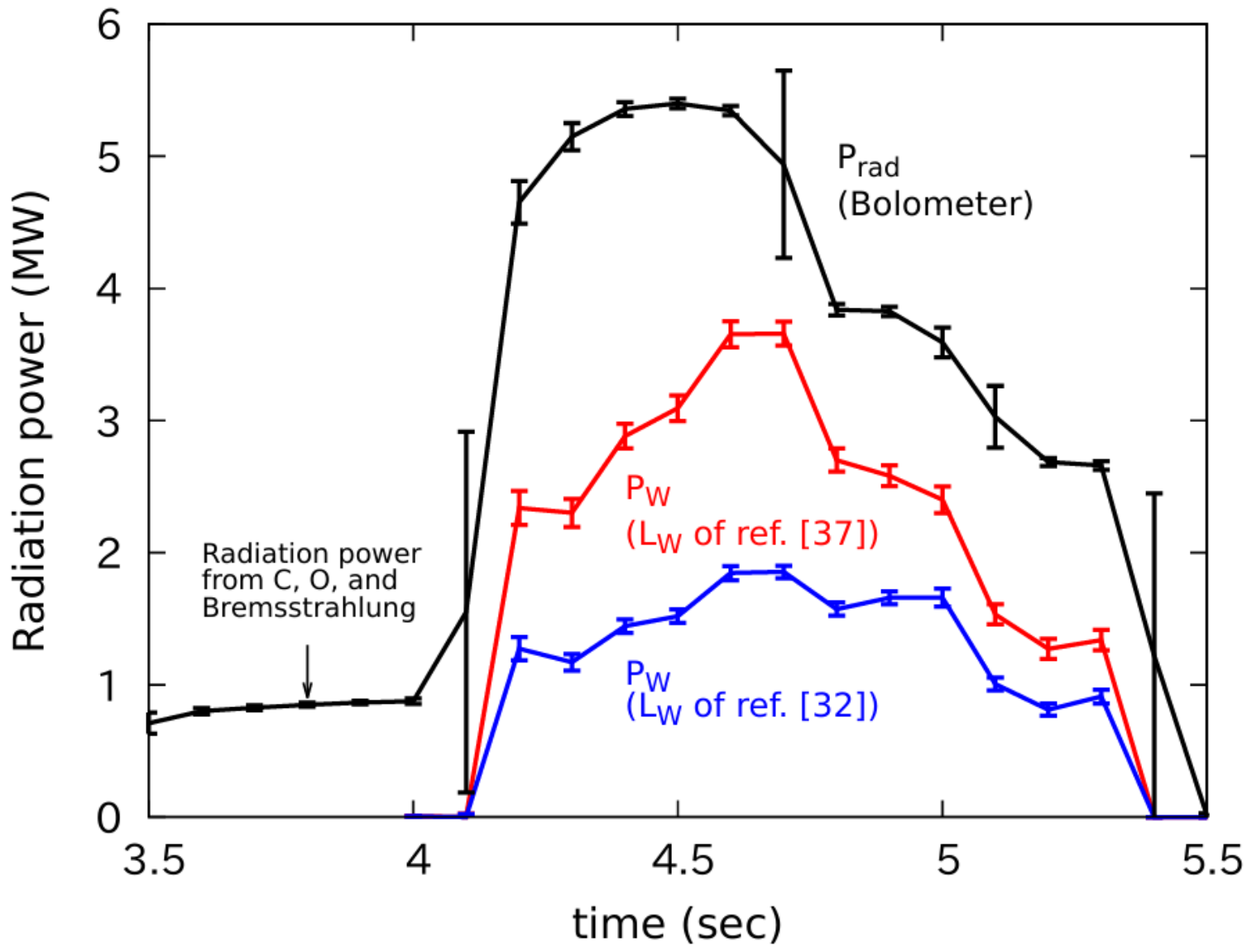

**Figure 13.** Temporal evolution of radiation power evaluated using cooling factor datasets provided by Ref. [37] and OPEN-ADAS [32], together with bolometer measurements.

Since the bolometer detects not only tungsten radiation but also radiation from other intrinsic impurities, such as C, O, as well as bremsstrahlung, the radiation power measured by the bolometer is higher than the $P_{\mathrm{W}}$ values estimated using the cooling factor datasets of Peyrusse *et al.* and OPEN-ADAS. Note also that, for $t$ < 4.65 s, the contribution of radiation power in $\rho$ > 0.9 is not included in the $P_{\mathrm{W}}$.

Despite these uncertainties, the $P_W$ estimated using the cooling factor dataset of Peyrusse *et al*. was in relatively good agreement with the radiation power measured by the bolometer. However, further investigation will be necessary in future studies as well as study of tungsten ions in charge states lower than $W^{17+}$ dominant in $\rho > 0.9$.

Bolometer measurements provide us with important insights into radiation power. Combined with the evaluation method for the spatiotemporal distribution of tungsten density presented in this paper, it will enable direct verification of the cooling factor dataset in future studies. In addition, bolometer arrays with multiple lines of sight are expected to be useful for reconstructing the spatiotemporal profile of radiation power through synthetic-diagnostic approaches such as the Slice & Stack method. The development of these techniques is likewise considered one of the important tasks.

## 4. Conclusion

Spectroscopic studies have been conducted in LHD with the pellet injection technique. Tungsten UTA spectra around 20-45, 45-70, 90-250, and 280-350 Å were clearly observed in a plasma with $T_e$ < 1 keV and $n_e = 10^{19}$-$10^{20}$ m$^{-3}$. Then, spatiotemporal profile of emissivity of 191.7±0.2 Å was analyzed using a space-resolved spectrometer, called EUV Long2.

The tungsten density profile was evaluated using photon emission coefficients, obtained using the CR model. In addition, radiation power was evaluated using the cooling factor datasets provided by OPEN-ADAS and by Peyrusse *et al.*

It was clearly observed that the injected pellet was first ablated at the edge plasma (ρ = 0.9). Subsequently, tungsten particles diffused throughout the plasma. After an event triggered by NBI breakdown, tungsten accumulated in the core plasma (ρ < 0.4). At least within the scope of this study, the radiation power, evaluated using the cooling factor dataset reported by Peyrusse *et al*., showed better agreement with the bolometer measurement, for plasmas with $T_e$ < 1 keV and $n_e > 10^{19}$-$10^{20}$ m$^{-3}$. By combining emissivity reconstruction methods such as Slice & Stack with atomic data, including PECs and cooling factors, more advanced impurity diagnostics, including tungsten density and radiation power profiles, can be achieved.

Further validation of atomic data, including ionization, recombination, excitation rate coefficients, PECs, and cooling factors, is required for tungsten ions in lower charge states. Even within the scope of this study, ions below $W^{16+}$, which are dominant at $T_e$ < 130 eV, were not included. These issues remain important tasks to be addressed for the steady-state operation of future fusion reactors.

## Funding

This work was partially supported by JSPS KAKENHI (Grant Numbers JP25KJ0583)

## Acknowledgements

Part of the experimental results in this research were obtained using supercomputing resources at Cyberscience Center, Tohoku University.

## Data availability

The raw data supporting the findings of this study are available in the LHD analyzed data repository at **https://doi.org/10.57451/lhd.analyzed-data** (accessed on 15 April 2026). The spectral data supporting the conclusions of this study will be made available by the authors upon request.

R. Nishimura : https://orcid.org/0009-0008-3408-9337 ryota.nishimura.b8[at]tohoku.ac.jp
T. Oishi : https://orcid.org/0000-0002-1171-8603 tetsutarou.oishi.a4[at]tohoku.ac.jp
I. Murakami : https://orcid.org/0000-0001-7544-1773 murakami.izumi[at]nifs.ac.jp
D. Kato : https://orcid.org/0000-0002-5302-073X kato.daiji[at]nifs.ac.jp
H.A. Sakaue : https://orcid.org/0000-0003-2209-3255 sakaue.hiroyuki[at]nifs.ac.jp
S. Gupta : https://orcid.org/0000-0002-7597-3240 gshivam475[at]gmail.com
H. Ohashi : https://orcid.org/0000-0001-5301-0526 ohashi[at]las.u-toyama.ac.jp
C. Suzuki : https://orcid.org/0000-0001-6536-9034 csuzuki[at]nifs.ac.jp
M. Goto : https://orcid.org/0000-0002-9160-682X goto.motoshi[at]nifs.ac.jp
Y. Kawamoto : https://orcid.org/0000-0002-6337-4080 kawamoto.yasuko[at]nifs.ac.jp
R.T. Ishikawa : https://orcid.org/0000-0002-4669-5376 ishikawa.ryohtaro[at]nifs.ac.jp
T. Kawate : https://orcid.org/0000-0002-1021-0322 kawate.tomoko[at]qst.go.jp
K. Mukai : https://orcid.org/0000-0003-1586-1084 mukai.kiyofumi[at]nifs.ac.jp
B.J. Peterson : https://orcid.org/0009-0002-8230-0121 peterson.byron[at]nifs.ac.jp
H. Takahashi : https://orcid.org/0000-0002-9700-6067 hiroyuki.takahashi.c6[at]tohoku.ac.jp